\documentclass[twocolumn,prl,superscriptaddress,showpacs,preprintnumbers,amsmath,amssymb]{revtex4}
\usepackage{graphicx}
\usepackage{dcolumn}
\usepackage{bm}

\begin{document}
\title{Geometrical Frustration: A Study of $4d$ Hard Spheres}
\author{J.A.~van~Meel}
\affiliation{FOM Institute for Atomic and Molecular Physics, Kruislaan 407, 1098 SJ Amsterdam, The Netherlands}
\author{D.~Frenkel}
\affiliation{Department of Chemistry, University of Cambridge, Lensfield Road, Cambridge CB2 1EW,UK}
\affiliation{FOM Institute for Atomic and Molecular Physics, Kruislaan 407, 1098 SJ Amsterdam, The Netherlands}
\author{P.~Charbonneau}
\affiliation{Department of Chemistry, Duke University, Durham, North Carolina 27708, USA}
\affiliation{FOM Institute for Atomic and Molecular Physics, Kruislaan 407, 1098 SJ Amsterdam, The Netherlands}
\date{\today}
\begin{abstract}
The smallest maximum kissing-number Voronoi polyhedron of $3d$ spheres is the icosahedron and the tetrahedron is the smallest volume that can show up in Delaunay tessalation.  No periodic lattice is consistent with either and hence these dense packings are geometrically frustrated.
Because icosahedra can be assembled from almost perfect tetrahedra, the terms ``icosahedral'' and ``polytetrahedral'' packing are often used interchangeably, which leaves the true origin of geometric frustration unclear.
Here we report a computational study of freezing of $4d$ hard spheres, where the
densest Voronoi cluster is compatible with the symmetry of the densest crystal, while polytetrahedral
order is not. We observe  that, under otherwise comparable conditions, crystal nucleation in $4d$ is  less facile than in $3d$.
This suggest that it is the geometrical frustration of polytetrahedral structures that inhibits crystallization.
\end{abstract}
\maketitle
Most glasses form under conditions where the thermodynamically
stable state of the system is crystalline. Good glass formers
should therefore be poor crystallizers. Geometrical frustration is
one of the factors that may prevent the formation of the ordered
phase and therefore help physical glass
formation~\cite{tarjus:2005}. There is also evidence that such
frustration increases the height of the crystallization-nucleation
barrier of liquid metals~\cite{schenk:2002}.
Isotropic simple liquids are often considered frustrated because the five-fold symmetry of the liquid icosahedron cannot pack as a regular lattice. This scenario contrasts with what happens in a fluid of $2d$ disks, where hexagonal order is both locally and globally preferred and where crystallization is particularly easy. 

Several physical mechanisms have been proposed to support the formation of icosahedra. On the one hand, Frank, considering the optimal way for kissing spheres to cluster around a central one, found the icosahedron to be more stable than the cubic lattice unit cells for the Lennard-Jones model~\cite{frank:1952}. Though the original argument relies on the energetics of spurious surface effects~\cite{doye:1996}, mean-field studies correcting for solvation leave the result unchanged~\cite{mossa:2003,mossa:2006}. The icosahedron, the smallest maximum kissing-number Voronoi polyhedron, is optimally packed. It offers the most free volume to surface spheres, so it is also preferred entropically. On the other hand, the polytetrahedral scenario ascribes the presence of icosahedra to their facile assembly from quasi-perfect tetrahedra, themselves the smallest Delaunay decomposition of space~\cite{nelson:2002,sadoc:1999}. But is it the packing of Voronoi polyhedra or the packing of Delaunay hyper-triangles that counts?
Experiments~\cite{dicicco:2003,aste:2005,dicicco:2007} and simulations~\cite{kondo:1991,jakse:2003} only manage to identify icosahedral order in limited quantities, even in deeply supercooled systems. 
Recent studies indicate that liquid polytetrahedral order is a lot more varied~\cite{anikeenko:2007,anikeenko:2008} than the icosahedral picture suggests.
Yet, because of the geometrical ambiguity, the equation of the icosahedron with frustration is difficult to asses.

Looking at crystallization in a system where polytetrahedral frustration does not correspond to a symmetric closed-shell structure like the icosahedron would help. Precisely such an
example is provided by the freezing of $4d$ spheres
that we study in this Rapid Communication.
It is, of course, somewhat unsatisfactory to perform a numerical
study of a system that cannot be probed experimentally. However,
there are other examples (e.g. renormalization-group theory) where
higher-dimensional model systems serve as a very useful reference
state for the theoretical description of our $3d$
world. The objective of the numerical study that we report here is
therefore not to present quantitative estimates of crystal
nucleation barriers in $4d$ (even though we obtain these numbers
too), but to shed more light on the nature and role of geometrical frustration and the ease of crystallization.

The $D_4$ crystal phase is formed by stacking, without voids,
24-cell Platonic polytopes~\cite{coxeter:1973,conway:1988}. In general, $D_d$ lattices are obtained by inserting an additional sphere in each
void of a $d$-dimensional cubic lattice. In $3d$ the spacing
between the spheres on the original cubic lattice increases
to form a body-centered-cubic crystal; in $4d$ the
additional sphere fits perfectly in the hole and leads to a unique, high symmetry crystal with maximal volume fraction $\eta=\pi^2/16\approx
0.617$. There exist other dense $4d$ lattices, such as
$A_4$ and $A_4^*$, but $D_4$ packs over $10\%$ more densely and offers more nearest-neighbor contacts. $D_4$'s unit cell, the 24-cell, is
made of 24 octahedral cells and is a Platonic polytope that has no analog in other dimensions~\cite{conway:1988}. Placing 24 kissing spheres around a central one in the 24-cell arrangement is
the densest closed-shell cluster of $4d$
spheres~\cite{musin:2004} and is postulated to be unique~\cite{pfender:2004}. Even accounting for solvation effects, clusters with the 24-cell geometry are locally preferred. Unlike in $3d$, for an equal number of particles $4d$ polytetrahedral clusters do not form more interparticle contacts than the 24-cell, and their slightly larger radius offers less, not more stabilization~\cite{mossa:2003}. The symmetry match between the 24-cell and the $D_4$ lattice therefore guarantees that no frustration arises from maximally kissing clusters. But neither the 24-cell nor any other unit cell can be assembled from (nearly) regular $4d$ tetrahedra. Four-dimensional spheres are thus an ideal system to clarify the origin of geometrical frustration. An earlier compaction study of $4d$ spheres indirectly hinted that spontaneous crystallization
might be slow~\cite{skoge:2006}, but this work could not
disentangle the different contributing factors, because such an analysis requires knowledge
of the equilibrium phase diagram, of the dynamical properties of
the fluid phase, and of the crystal nucleation barriers.  Our
computational study addresses these questions. To this end, we first locate the $4d$ freezing transition, quantify the fluid order, and then compute the free energy barrier to nucleation at different
supersaturations.

\begin{figure}
\center{\includegraphics[width=0.9\columnwidth]{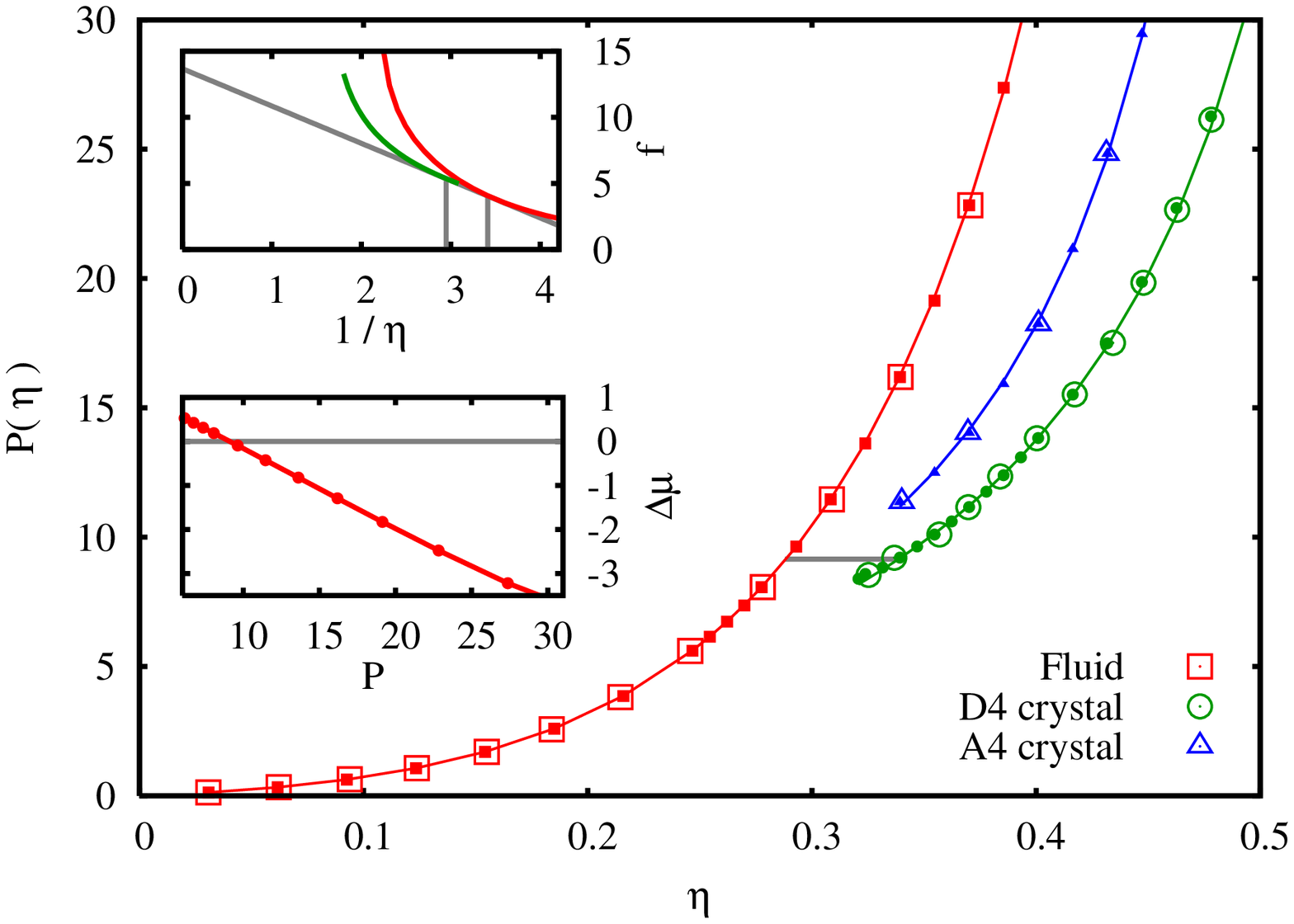}}
\caption[Phase diagram]{(Color online) Equations of state of 
$4d$ hard spheres
at constant $V$ (closed) and $P$ (open) extend earlier molecular dynamics results~\cite{michels:1984}. Pad\'e approximant
for the fluid~\cite{bishop:2005} and Speedy fits for the
crystals~\cite{speedy:1998} are given for reference (lines). 
Bottom inset: at coexistence chemical potentials are equal, 
thus $P_{\mathrm{coex}}=9.15$ and $\mu_{\mathrm{coex}}=13.68$.
Top inset: the common tangent to the free energy curves
pinpoints the phase transition boundaries:
$\eta_{\mathrm{freeze}}=0.288$ and $\eta_{\mathrm{melt}}=0.337$.}
\label{fig:phasediagram}
\end{figure}
Interestingly, although the
equations of state of both the fluid and the crystal phases of $4d$
hard spheres were computed in the early 80's~\cite{michels:1984},
we are not aware of any numerical determinations of the solid-fluid
coexistence point. Using a quasi-Maxwell
construction~\cite{streett:1974} at the crystal stability limit~\cite{skoge:2006}, we can use these results to
approximate the coexistence range $\eta_{\mathrm{coex}}=0.29 - 0.35$,
but this is insufficiently accurate. To the best of our knowledge, density functional theory has only been applied to the fluid-$A_4$
coexistence~\cite{colot:1986}. In order to precisely locate the freezing point, we thus performed standard
$NVT$-Monte Carlo (MC) simulations to compute the equation of state of hard
spheres, outside the range studied in
Ref.~\cite{michels:1984}. As a test, we also performed constant
$NPT$ simulations and verified that the two techniques yielded
consistent results. In what follows, we use the particle diameter
$\sigma$ as our unit of length and the thermal energy $k_BT$
as our unit of energy.  The equation of state of $4d$ spheres is
related to the value of the pair-distribution function $g(r)$ at
contact
$Pv_0/\eta= 1+8\eta g(1^+)$,
where $v_0$ is the volume of a
$4d$ sphere and $g(1^+)$ is the value of the radial distribution
function at contact~\cite{bishop:2005}. The results for the fluid
and two crystal phases are presented in
Fig.~\ref{fig:phasediagram} for systems containing $2048$ ($D_4$)
and $4096$ (fluid and $A_4$) particles. The equation of state
could not be calculated for $A_4^*$, because it is
mechanically unstable, which makes it unlikely to contribute to the
crystallization process. We won't consider it further. To locate the fluid-solid coexistence regime, we need
to determine the absolute free energy of the solid at least at
one point~\cite{frenkel:2002}.
The absolute Helmholtz free energy
per particle $f$ of the $D_4$ and $A_4$ crystals at $\eta=0.37$ is
obtained by the Einstein-crystal method~\cite{frenkel:1984}. The
free energy at other densities can then be obtained by thermodynamic integration.
We find $D_4$ to be the thermodynamically stable crystal phase. The
fluid-$D_4$ coexistence pressure $P_{\mathrm{coex}}$
(Fig.~\ref{fig:phasediagram} lower inset), allows to read off
the melting and freezing densities by common tangent construction
(Fig.~\ref{fig:phasediagram} higher inset). The resulting two-phase
region $\eta_{\mathrm{coex}}=0.288-0.337$ is compatible with the rough estimate above.
The thermodynamic driving force for crystallization in the
supersaturated fluid at constant pressure is the difference in
chemical potential $\Delta\mu=\mu_{D_4}-\mu_{\mathrm{fluid}}$
between the two phases displayed in
the lower inset of Fig.~\ref{fig:phasediagram} .

\begin{figure}
\center{\includegraphics[width=0.9\columnwidth]{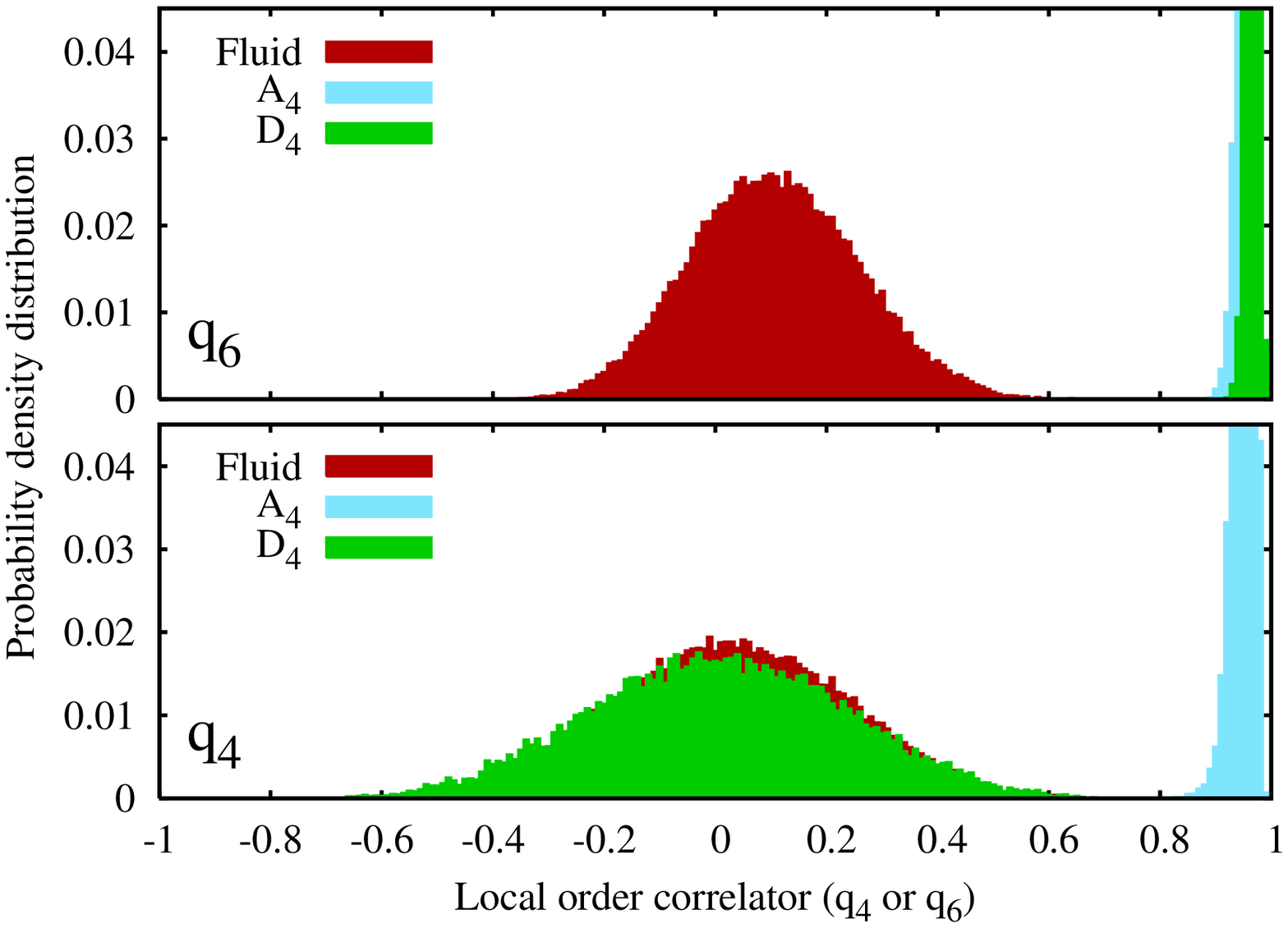}}
\caption[Order parameter]{(Color online) Distribution of the local order
correlator 
with $l=4$ (top) and $l=6$ (bottom) 
at $P=19$.}
\label{fig:orderparameter}
\end{figure}
To characterize the structure of the fluid and identify
the formation of crystallites, we need a local criterion that distinguishes crystal from fluid.
Studies in $2d$ and $3d$ suggest that order
parameters derived from invariant combinations of spherical harmonics
$Y_l$ of degree $l$ might suffice~\cite{tenwolde:1996,auer:2001}. In high dimensions, it is more convenient to rewrite
the second-order invariant in terms of Gegenbauer
polynomials $G_l^n$, where $n=d/2-1$, using the sum rule~\cite{pfender:2004}. The $(l+1)^2$ $4d$ spherical harmonics give
\begin{equation}
G_l^1(\hat{\mathbf{r}}_1\cdot\hat{\mathbf{r}}_2)=\frac{2\pi^2}{(l+1)^2} \sum_{m=1}^{(l+1)^2}Y_l^m (\hat{\mathbf{r}}_1) \overline{Y_l^{m} (\hat{\mathbf{r}}_2)},
\end{equation}
where $\hat{\mathbf{r}}_i$ are unit vectors. The local order correlator is
\begin{equation}
q_l^{i,j}=\mathbf{q}_l(i)\cdot\mathbf{q}_l(j)=\frac{1}{N(i)N(j)} \sum_{\alpha=1}^{N(i)}\sum_{\beta=1}^{N(j)}G_l^1 (\hat{\mathbf{r}}_{\alpha i}\cdot\hat{\mathbf{r}}_{\beta j}), \label{eq:ord_param}
\end{equation}
where the indices $\alpha$ and $\beta$ run over the number of
neighbors contained within
a distance equal to the first minimum of $g(r)$.
The local order correlation
distinguishes between different geometrical environments: $q_6$ set apart fluidlike particles from those within a $D_4$ or
an $A_4$ lattice, while $q_4$ discriminates between the two crystals (Fig.~\ref{fig:orderparameter}). 

\begin{figure}
\center{\includegraphics[width=0.9\columnwidth]{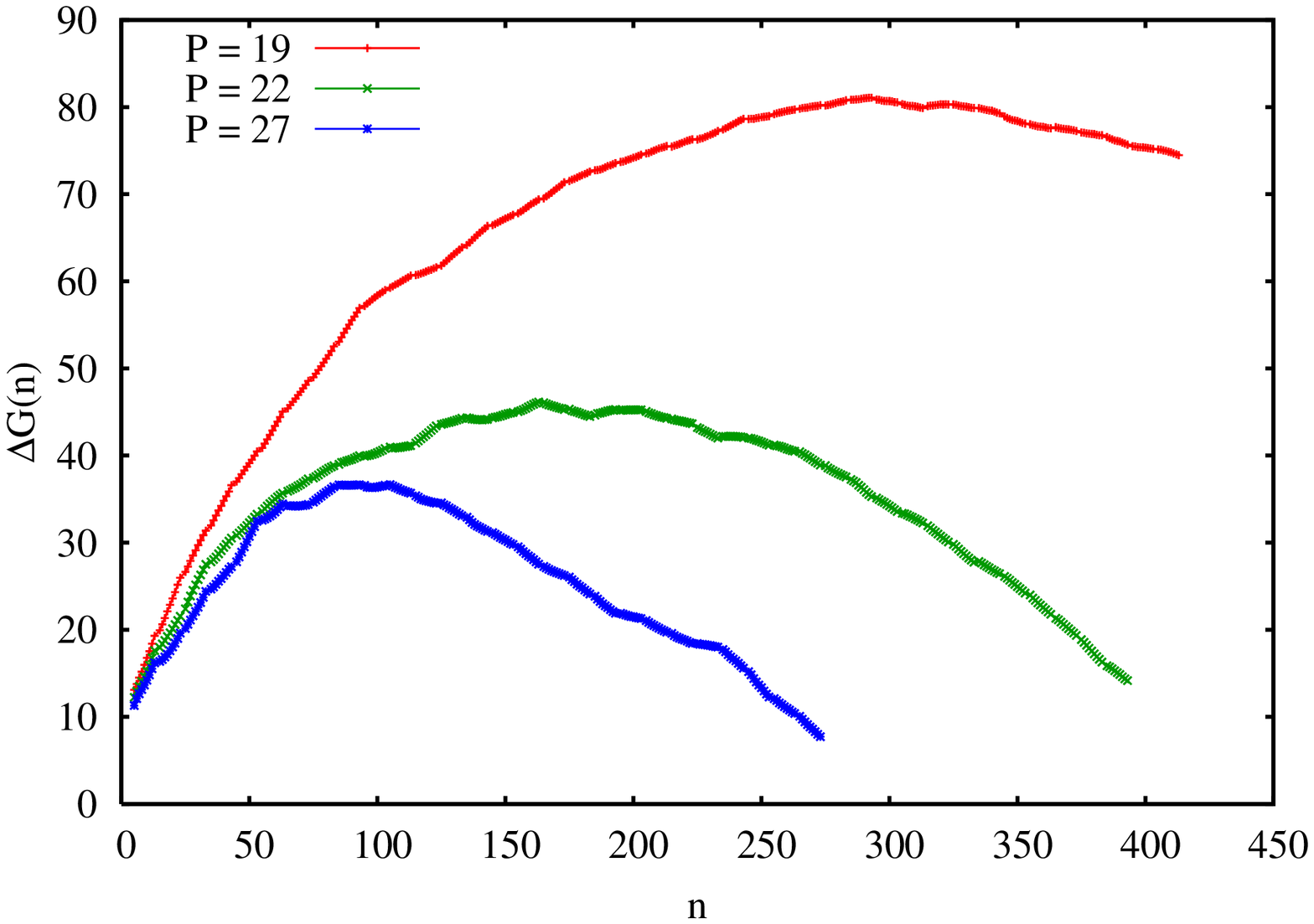}}
\begin{ruledtabular}
  \begin{tabular}{|c|c|c|c|c|c|}
  $P$ & $\Delta\mu$ & $\Delta G^*$ &$\gamma_{\mathrm{CNT}}$ & $n^*$ &
  $n^*_{\mathrm{CNT}}$\\
  \hline
  $19$ &  $-1.8$ & $81$ & $1.80$ & $157$ & $133$\\
  $22$ &  $-2.3$ & $42$ & $1.94$ & $75$  & $60$ \\
  $27$ &  $-3.2$ & $37$ & $2.4 $ & $40$  & $35$ \\
\end{tabular}
\end{ruledtabular}
\caption[Crystallization barrier]{(Color online) Free energy
barriers for $4096$ particles at various supersaturations, along with the corresponding CNT and simulation parameters.
The critical cluster size $n^*$ is obtained using a stricter
order parameter (see text).
} \label{fig:cnt}
\end{figure}

As freezing in $4d$ is a first-order phase transition, we expect crystallization to proceed via nucleation and growth. A Landau free energy analysis predicts that crystals with reciprocal lattice vectors forming equilateral triangles should initiate the nucleation~\cite{alexander:1978}. Though this argument has met only limited success in $3d$~\cite{auer:2001}, in $4d$ it supports the preferential nucleation of $D_4$, in line with the thermodynamic drive.
To estimate the ease of crystallization, we compute the free energy barrier for crystal nucleation $\Delta
G^*$. Classical  nucleation theory (CNT)~\cite{volmer:1926} derives from the
thermodynamic drive $\Delta \mu$ and the interfacial free energy
$\gamma$ of a spherical crystallite a free energy
functional that depends on the size $n$ of the crystallite
\begin{equation}
\Delta G(n)=S_d (n/\rho_{\mathrm{x}})^{(d-1)/d} \gamma -n\Delta\mu,
\label{eq:cnt}
\end{equation}
where $\rho_{\mathrm{x}}$ is the crystal density at a given pressure and the shape-dependent prefactor is $S_4=(128\pi^2)^{1/4}$ for $4d$ spheres. The resulting maximal barrier height is then
\begin{equation}
\Delta G^*(n^*)=\frac{27\pi^2\gamma^4}{2\rho_{D_4}^3\Delta\mu^3}
\label{eq:barrheight}
\end{equation}
at the critical cluster size $n^*$. The rate of nucleation per unit volume $I$ is given by $I=\kappa \exp(-\Delta G^*)$, where $\kappa$ is a kinetic prefactor that is proportional to the diffusion coefficient in the fluid phase~\cite{auer:2001}.
Though schematic this level of theory is sufficient to clarify the
contribution of geometrical frustration through an analysis of
$\gamma$. Within the CNT framework the geometrical mismatch in
$3d$ between icosahedral and crystal order should lead to a
relatively large $\gamma$, while in $4d$ one might expect $\gamma$ to be small if the locally preferred cluster scenario is valid,
but not for polytetrahedral frustration. 

Results for $3d$ crystallization are
available~\cite{auer:2001},  so only a few $4d$
barriers are needed to complete the picture. Crystallization being a
rare event in this regime, we perform constant-pressure MC runs
with umbrella sampling to bias the growth of
a crystal cluster from the fluid~\cite{frenkel:2002}. A standard algorithm is employed to identify the crystallites~\cite{tenwolde:1996,auer:2001}. We \emph{link}
pairs of nearest neighbors with $q_6>0.4$.
If a particle has more than
five links it is deemed crystalline. The number of
particles in the largest crystallite is then the order
parameter. The resulting free energy profiles are presented in
Fig.~\ref{fig:cnt}. Though $q_6$ does not
discriminate between $A_4$ and $D_4$ crystals, further
checks with $q_4$ show that only the latter nucleates. In $4d$ a low $q_6$ cutoff value is
required, because of the minimal overlap between the crystalline and fluid
regions (Fig.~\ref{fig:orderparameter}), and consequently, non-compact
clusters are initially observed. Though the clusters irreversibly compactify, the process
can be very slow. To reduce the computational burden, the system is first equilibrated
by growing the total number of \emph{links} in the largest crystallite.
The low $q_6$ cutoff also artificially inflates the measured critical cluster size.
A fit to the CNT functional form (Eq.~\ref{eq:cnt}) is thus of little use in extracting
$\gamma$. However, because the barrier height is unaffected by this biasing choice
and $\Delta \mu$ is known, we can obtain
$\gamma$ from Eq.~\ref{eq:barrheight} directly. To validate the implied size of the CNT critical
cluster $n^*_{\mathrm{CNT}}$ we compare it to the cluster
size obtained by imposing a purely crystalline linking criterion $q_6>0.65$
to the configurations at the top of the barrier.
The difference between the two (Fig.~\ref{fig:cnt})
is no more than $25\%$, which is remarkably good in this context.

The results of Fig.~\ref{fig:cnt} allow us to conclude that the very slow crystallization of $4d$ spheres observed in the study of Ref.~\cite{skoge:2006} is due to the presence of a considerably higher $4d$ nucleation barrier than at the same supersaturation in $3d$.
Slow nucleation could also be due to a low value of the kinetic prefactor $\kappa$,
which would require that the diffusion of particles in the dense fluid be anomalously slow. But simulations with the code of Ref.~\cite{skoge:2006} show no evidence for slow
diffusion, not even at the highest pressures studied. The slow crystallization is thus consistent with a high degree of geometrical frustration in $4d$ fluids. Based on the similarity between the number of neighbors within the first peak of $g(r)$ and the maximal kissing number Skoge~\emph{et al.} speculated that high dimensional fluids contain a number of deformed crystalline unit cells, rather than polytetrahedral structures~\cite{skoge:2006}. However, the clear difference between the fluid and the 24-cell shown by the local order correlator (Fig~\ref{fig:orderparameter}) suggests this not to be the case. The similarity between the kissing number and the number of first neighbors can instead be explained by a wide first peak of $g(r)$ (not shown) that accommodates non-kissing neighbors in polytetrahedral clusters. Because the ``locally preferred'' 24-cell has little to do with geometrical frustration, our results support the generic polytetrahedral structures as the source of frustration.
By dimensional analogy, we infer that the ``locally preferred'' icosahedron is not singular, but instead one of the many possible geometrically frustrating structures, and explains its limited presence in fluids. The dimensionless surface free-energy density $\gamma\sigma^{d-1}/(k_BT)$ is at least two to three times larger in $4d$ than in $3d$, which indicates that geometrical frustration is surprisingly rather weak in $3d$. It is this weakness that helps make hard sphere crystallization so prevalent. The interesting puzzle is therefore not to identify the origin of $3d$ frustration, but the source of its mildness. One possibility is that the tetrahedra that are part of the face-centered cubic (fcc) structure (none are found in $D_4$) relax the geometrical frustration and therefore reduce the interfacial tension. Another possibility is that the ``planetary perturbations'' that allow to exchange the positions of spheres at the surface of an icosahedron by sliding, go through a cuboctahedron configuration, which is the fcc unit cell~\cite{conway:1988}. If common, this phenomenon would imply that not all polytetrahedral structures are equally frustrating and that icosahedra might in fact be early nucleation sites.

The large values for the height of the nucleation barrier of $4d$
crystals, as well as the evidence (Fig.~\ref{fig:orderparameter}) that the local structures
in the fluid and the $D_4$ crystal are rather different, indicate that it is the Delaunay packing that matters. This finding underlines that
one should be rather careful in caricaturing the nature of frustration as icosahedral
in $3d$ liquids. Icosahedra are but one of the many possible polytetrahedral arrangements and little indicates that it plays a more prominent role in geometrical frustration than others.
Note that the difficulty to crystallize $4d$ spheres makes them, as well as their higher dimensional equivalents, promising testing grounds for theories of packing and glass-forming liquids.

We thank B.~Charbonneau, G. Tarjus, R. Mossery, B.~Mulder and S.~Abeln for their help at various stages of this project.
The work of the FOM Institute is part of
the research program of FOM and is made possible by financial
support from the Netherlands Organization for Scientific Research
(NWO).
\bibliography{text}
\end{document}